\documentclass[11pt]{article}
\usepackage{graphicx}

\newcommand{\BABARPubYear}    {03}

\newcommand{\BABARConfNumber} {019}
\newcommand{\SLACPubNumber} {10100}

\input babarsym

\def\BtoXsll     {\ensuremath{\B \to X_s\: \ell^+ \ell^-}}
\def\BtoXsee     {\ensuremath{\B \to X_s\: e^+ e^-}}
\def\BtoXsmumu   {\ensuremath{\B \to X_s\: \mu^+ \mu^-}}
\def\BtoXsemu    {\ensuremath{\B \to X_s\: e^\pm \mu^\mp}}
\def\BtoXshh     {\ensuremath{\B \to X_s\: \pi^+ \pi^-}}
\def\BtoKll      {\ensuremath{\B \to K \ell^+ \ell^-}}
\def\BtoKstarll  {\ensuremath{\B \to K^\ast \ell^+ \ell^-}}
\def\BtoJpsiX    {\ensuremath{\B \to J/\psi\:X}}
\def\BtopsipX    {\ensuremath{\B \to \psi(2S)\:X}}

\long\def\inst#1{\par\nobreak\kern 4pt\nobreak
    {\it #1}\par\vskip 10pt plus 3pt minus 3pt}

\begin{document}
{\pagestyle{empty}

\begin{flushright}
\babar-CONF-\BABARPubYear/\BABARConfNumber \\
SLAC-PUB-\SLACPubNumber \\
August 2003 \\
\end{flushright}

\par\vskip 5cm

% Title of the paper
\begin{center}
\Large \bf Measurement of the \BtoXsll\ Branching Fraction \\
           Using a Sum over Exclusive Modes
\end{center}
\bigskip

\begin{center}
\large The \babar\ Collaboration\\
\mbox{ }\\
\today
\end{center}
\bigskip \bigskip

% Abstract
\begin{center}
\large \bf Abstract
\end{center}

We present a measurement of the branching fraction for the
flavor-changing neutral current process \BtoXsll\ based on
a sample of $88.9 \times 10^{6}$~$\Upsilon(4S) \to \BB$ events 
recorded with the \babar\ detector at the \pep2\ $e^{+}e^{-}$ storage ring.
The final state is reconstructed from pairs of electrons or muons
and a hadronic system consisting of one $K^{\pm}$ or $K^{0}_{s}$ and up to 
two pions, with at most one $\pi^{0}$. 
Summing over both lepton flavors, we observe
a signal of $41 \pm 10(stat) \pm 2(syst)$ events
with a statistical significance of $4.6\:\sigma$. 
The inclusive branching fraction is determined to be
${\cal B}(\BtoXsll) = 
 \left(6.3 \pm 1.6(stat) ^{+1.8}_{-1.5}(syst)\right) \times 10^{-6}$
for $m(\ell^{+} \ell^{-}) > 0.2$~\gevcc.
All results are preliminary.

\vfill
\begin{center}
Contributed to the XXI$^{\rm st}$ International Symposium
on Lepton and Photon Interactions at High~Energies, 8/11 --- 8/16/2003, Fermilab, Illinois, USA.
\end{center}

\vspace{1.0cm}
\begin{center}
{\em Stanford Linear Accelerator Center, Stanford University, 
Stanford, CA 94309} \\ \vspace{0.1cm}\hrule\vspace{0.1cm}
Work supported in part by Department of Energy contract DE-AC03-76SF00515.
\end{center}

\newpage
} % end of pagestyle{empty}

% Input author list file
\begin{center}
\small

The \babar\ Collaboration,
\bigskip

%% author list as of 02-Jun-2003 (595 authors)
%
B.~Aubert,
R.~Barate,
D.~Boutigny,
J.-M.~Gaillard,
A.~Hicheur,
Y.~Karyotakis,
J.~P.~Lees,
P.~Robbe,
V.~Tisserand,
A.~Zghiche
\inst{Laboratoire de Physique des Particules, F-74941 Annecy-le-Vieux, France }
% ...

\end{center}\newpage

\section{Introduction}
\label{sec:Introduction}

\begin{figure}[h]
\begin{center}
\includegraphics[width=14cm]{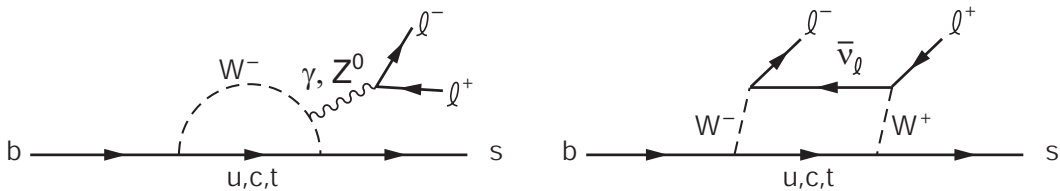}
\caption{Feynman diagrams for $b \to s \ell^{+} \ell^{-}$ transitions.}
\label{fig:feynman}
\end{center}
\end{figure}

%%% general intro
The rare decay \BtoXsll\ proceeds through a 
$b \to s \ell^{+} \ell^{-}$ transition, which is forbidden at tree 
level in the Standard Model (SM). 
However, such a flavor-changing neutral current (FCNC)
process can occur at higher order via electroweak penguin and $W^{+} W^{-}$
box diagrams, as shown in Figure~\ref{fig:feynman}.
The $b \to s \ell^{+} \ell^{-}$ transition therefore allows 
deeper insight into the effective Hamiltonian describing FCNC processes
and is sensitive to the effects of non-SM physics that may enter these
loops; see, for example, Refs.~\cite{Ali02,Hurth03}.

Recent calculations of the branching fractions 
for the exclusive decay modes predict
${\cal B}(B \to K e^{+} e^{-}) = {\cal B}(B \to K \mu^{+} \mu^{-}) = 
(0.35 \pm 0.12) \times 10^{-6}$,
${\cal B}(B \to K^{*} e^{+} e^{-}) = (1.58 \pm 0.49) \times 10^{-6}$
and 
${\cal B}(B \to K^{*} \mu^{+} \mu^{-}) = (1.19 \pm 0.39) \times 10^{-6}$,
and for the inclusive processes
${\cal B}(\BtoXsee) = (6.9 \pm 1.0) \times 10^{-6}$
and
${\cal B}(\BtoXsmumu) = (4.2 \pm 0.7) \times 10^{-6}$~\cite{Ali02}.
In the electron channel, the branching fraction is predicted 
to be ${\cal B}(\BtoXsee) = (4.2 \pm 0.7) \times 10^{-6}$~\cite{Ali_ICHEP02}
for $m(e^{+} e^{-}) > 0.2$~\gevcc.
Measurements of the inclusive branching fractions are 
motivated by the smaller theoretical uncertainties, as compared with 
predictions for exclusive decays.
Both the Belle and \babar\ Collaborations have 
observed the $B \to K \ell^{+} \ell^{-}$ ($\ell = e, \mu$)
decay~\cite{Belle_Kll,BaBar_Kll},
\babar\ has found evidence for the \BtoKstarll\ decay~\cite{Ryd},
and Belle has measured the inclusive \BtoXsll\ decay~\cite{Belle_sll}.

In the present analysis, we study the \BtoXsll\
process by reconstructing the final state from pairs of electrons
or muons and a hadronic system consisting of one $K^{\pm}$ or $K^{0}_{s}$ 
and up to two pions, with at most one $\pi^{0}$. 
The choice of maximum number of pions is motivated by the fact that
the signal-to-background ratio drops significantly with increasing
multiplicity.
This approach~\cite{CLEO_bsgamma95}, which sums over exclusive modes,
allows approximately half of the full inclusive rate to be reconstructed.
If the fraction of modes containing a \KL\ 
is taken to be equal to that containing a \KS,
the missing states that remain unaccounted for represent $\sim$25\%
of the total rate.
We require the hadronic mass for the selected final states
to be less than 1.8~\gevcc\ to reduce combinatorial background. 
This cut retains approximately 94\% of the signal in the reconstructed modes.  
We correct for the missing modes and the effect of the 
hadronic mass cut to extract the 
inclusive \BtoXsll\ decay rate for $m(\ell^{+} \ell^{-}) > 0.2$~\gevcc.

\section{The \babar\ Detector and Data Sample}
\label{sec:babar}

%%% data sample
The data sample used in this analysis was collected with the \babar\ detector
during 
1999-2002 at the \pep2\ asymmetric-energy $e^{+}e^{-}$ storage ring
at the Stanford Linear Accelerator Center.
It consists of $81.9$~\invfb\ recorded
at the $\Upsilon(4S)$ resonance, corresponding to 
$88.9 \times 10^{6}$~$\Upsilon(4S) \to \BB$ events,
as well as an additional $9.6$~\invfb\ recorded $40$~\mev\ below 
the $\Upsilon(4S)$ resonance. The off-resonance data provide
a clean sample of $e^{+}e^{-} \to q\bar{q}$ (with $q = u,d,s,c$) 
continuum events for background studies.

%%% detector
This analysis relies on the charged-particle tracking systems and particle 
identification systems of the \babar\ detector~\cite{ref:babar}.
A 5-layer silicon vertex tracker (SVT) and a 40-layer drift chamber (DCH) 
provide tracking and particle identification for charged particles.
The DIRC, a Cherenkov ring-imaging system, is primarily used for 
charged hadron ($\pi, K, p$) identification.
The electromagnetic calorimeter (EMC),
consisting of Tl-doped CsI crystals, provides electron identification
and photon reconstruction.
These detectors are located inside a 1.5-T solenoidal superconducting
magnet. Muons are identified in the instrumented flux return (IFR)
with resistive plate chambers interleaved with iron plates.

%%% particle selections, brem recovery for electrons
Clean particle identification for the final state particles 
$e^{\pm}$, $\mu^{\pm}$, $K^{\pm}$, $K^{0}_{s}$, $\pi^{\pm}$ and $\pi^{0}$
is important for this analysis. 
Electrons and muons are required to have lab-frame momenta
greater than $0.5$~\gevc\ and $1.0$~\gevc, respectively. 
Bremsstrahlung photons from electrons are recovered by
combining an electron with up to three photons
within a small angular region around the electron direction~\cite{bremrecovery}.
$K^{0}_{s}$ candidates are reconstructed from pairs 
of oppositely-charged tracks with 
$|m(\pi^{+}\pi^{-}) - m(K^{0}_{s})| < 11.2$~\mevcc,
decay length greater than $2$~mm, and angle between
the $K^{0}_{s}$ momentum vector and the line between the primary vertex 
and the $K^{0}_{s}$ vertex satisfying $\cos\delta > 0.99$.
Charged pions are selected from tracks after rejection of 
cleanly identified $e^{\pm}$ and $K^{\pm}$ candidates. 
Neutral pions are required to have lab-frame energy greater than $400$~\mev, 
photon daughter energies greater than $50$~\mev,
and a $\gamma \gamma$ invariant mass that satisfies 
$|m(\gamma \gamma) - m(\pi^{0})| < 10$~\mevcc.

\section{Analysis overview}
\label{sec:overview}

%%% analysis approach, concerns which drive analysis
We use a technique that sums over exclusive modes to measure the inclusive
branching fraction ${\cal B}(\BtoXsll)$.
Compared to a fully inclusive approach, this method has the advantage
of being able to exploit the strong kinematical discrimination provided by 
$m_{ES} = \sqrt{E_{beam}^2 - \vec{p}_{B}^{\,2}}$
and $\Delta E = E_{B} - E_{beam}$,
where $E_{beam}$ is the beam energy
and $E_{B}$ ($\vec{p}_{B}$) is the 
reconstructed $B$ meson energy (3-momentum).
All quantities are evaluated in the $e^{+}e^{-}$ center-of-mass system (CM). 
This analysis approach is needed to suppress large backgrounds from
\BB\ and continuum events, and to extract the expected small signal yield.

The main contribution to the combinatorial background is
from semileptonic decays in \BB\ events.
In these events, \BtoXsll\ candidates contain decay products from
both $B$ mesons (i.e., the daughters of the \BtoXsll\ candidate do not 
originate from the same point) and there is a significant amount of missing energy
due to neutrinos.
Another contribution to the combinatorial background is due
to continuum events, which are effectively suppressed with event-shape variables.

The dominant peaking backgrounds, which are from 
$B \to J/\psi X$ and $B \to \psi(2S) X$ decays with
$J/\psi (\psi(2S)) \to \ell^{+} \ell^{-}$, 
mimic the signal and need to be efficiently removed with cuts on the 
dilepton mass $m(\ell^{+} \ell^{-})$. The resulting veto sample provides
a large control sample of decays with a signature identical to that
of the signal, albeit in a restricted range of $m(\ell^{+} \ell^{-})$.
$B \to D^{(*)} n \pi$~$(n > 0)$ decays with misidentification 
of two charged pions as leptons present an additional source of peaking background.

%%% MC generator
Simulated \BtoXsll\ events are produced with a combination of
exclusive and inclusive models.
In the hadronic mass region of $m(X_{s}) < 1.1$~\gevcc, 
exclusive $B \to K^{(*)} \ell^{+} \ell^{-}$ decays are generated 
according to~\cite{Ali02,Ali00},
where the relevant form factors are computed using light-cone
QCD sum rules. The remaining decays, in the region $m(X_{s}) > 1.1$~\gevcc, 
are generated with a nonresonant model following~\cite{Ali02,Kruger96} 
and using the Fermi motion model of~\cite{Ali79}. 
JETSET~\cite{JETSET} is then used to hadronize 
the system consisting of a strange quark and a spectator quark.

\section{Event selection}
\label{sec:selection}

%%% skim
Events are required to have a well determined primary vertex,
be tagged as multi-hadron events,
have a ratio of the second to zeroth-order Fox-Wolfram
moments~\cite{FoxWolf}
(calculated using charged tracks and neutral calorimeter clusters)
of $R_{2} < 0.5$, have at least four charged tracks, and contain 
at least two leptons (either electrons or muons) satisfying loose
particle identification criteria and having lab-frame momenta greater than 
$0.5$~\gevc\ for electrons and $0.8$~\gevc\ for muons.

%%% B candidate selection   
In the next step, $B$ candidates are reconstructed.
A dilepton is selected by choosing the $e^{+} e^{-}$ or
$\mu^{+} \mu^{-}$ pair with the largest value of 
$|p(\ell^{+})| + |p(\ell^{-})|$, using lab-frame momenta.
The dilepton candidate is required to form a good vertex.
Using this $\ell^{+} \ell^{-}$ pair,
$B \to X_{s} \ell^{+} \ell^{-}$ candidates are formed 
by adding either a $K^{\pm}$ or a $K^{0}_{s}$ and up to
two pions, but no more than one $\pi^{0}$. 
In this manner, ten different hadronic topologies
are considered:
$K^+$, $K^+ \pi^0$, $K^+ \pi^-$, $K^+ \pi^- \pi^0$, $K^+ \pi^- \pi^+$,
$\KS$, $\KS \pi^0$, $\KS \pi^+$, $\KS \pi^+ \pi^0$, and $\KS \pi^+ \pi^-$.

%%% preselection
The resulting $B $~candidates are required to have an
$\ell^{+} \ell^{-}$ vertex fit probability $\log(P_{ll vtx}) > -10$,
a $B$ vertex constructed from charged daughter particles 
with fit probability $\log(P_{B vtx}) > -10$,
$m(X_{s}) < 2.5$~\gevcc, $5.0 < m_{ES} < 5.29$~\gevcc, and $|\Delta E| < 0.3$~\gev.

%%% pick at most 1 B cand per event
At this stage, there is an average of five $B$ candidates 
per event in the signal simulation. 
Only the $B$~candidate with the largest signal likelihood 
value is retained.
The signal likelihood function is constructed based on the simulated 
distributions of $\Delta E$, $\log(P_{B vtx})$, and $\cos\theta_{B}$
for the signal, where $\theta_{B}$ is the angle between the momentum vector 
of the $B$~candidate and the beam axis in the CM frame.
We select the best candidate before further selection cuts are applied.
This approach entails a small cost in signal efficiency,
which is more than compensated by an improved background suppression
in the final sample.

%%% bkg suppression

%%% postselection
After selecting the best candidate, combinatorial backgrounds
are suppressed by tightening some of the earlier requirements:
$m(X_{s}) < 1.8$~\gevcc,
$5.20 < m_{ES} < 5.29$~\gevcc,
and $-0.2 < \Delta E < 0.1$~\gev. 
In addition, we require the separation between the two leptons 
along the beam direction to satisfy $|\Delta z| < 0.15$~cm,
where the $z$-coordinate of each lepton is determined at 
the point of closest approach to the beam axis.
We also require the rest of the event, formed by combining all charged tracks
and neutral calorimeter clusters not included in the $B$~candidate,
to satisfy $-5.0 < \Delta E^{ROE} < 2.0$~\gev\ and $m^{ROE}_{ES} > 4.9$~\gevcc.

%%% cveto
Charmonium backgrounds are reduced by
removing $B$~candidates with dilepton mass in the ranges
$2.70 < m(e^{+} e^{-}) < 3.25$~\gevcc,
$2.80 < m(\mu^{+} \mu^{-}) < 3.20$~\gevcc,
$3.45 < m(e^{+} e^{-}) < 3.80$~\gevcc, and
$3.55 < m(\mu^{+} \mu^{-}) < 3.80$~\gevcc.
If one of the electrons from a $J/\psi$ or $\psi(2S)$ decay erroneously 
picks up a random photon in the Bremsstrahlung-recovery process,
the dilepton mass can increase sufficiently to evade
the above cuts.
Therefore the charmonium veto is applied to the dilepton mass 
before and after Bremsstrahlung recovery.
Using the simulation, we estimate the remaining peaking 
charmonium background to be $0.6 \pm 0.3$ events and
$0.3 \pm 0.2$ events for $e^{+} e^{-}$ modes and $\mu^{+} \mu^{-}$
modes, respectively.

%%% B -> Xs gamma
The potential peaking background from $B \to X_{s} \gamma$ decays,
followed by conversion of the photon into an $e^{+} e^{-}$ pair
in the detector material, is a concern for the $e^{+} e^{-}$ modes 
only. After observing an excess of events in data over the simulation for
low $e^{+}e^{-}$ mass in the $m_{ES} < 5.27$~\gevcc\ sideband
distribution, we completely remove this background by requiring
$m(e^{+}e^{-}) > 0.2$~\gevcc. 

%%% LR (9 vars)
The final suppression of the combinatorial background is
achieved with a likelihood ratio ${\cal L}_{R}$ based on nine variables:
$\Delta E$, $\Delta E^{ROE}$, $m^{ROE}_{ES}$, $\Delta z$,
$\log(P_{B vtx})$, $\cos\theta_{miss}$, where $\theta_{miss}$ 
is the angle between the missing momentum vector for
the whole event and the $z$~axis in the CM frame, $\cos\theta_{B}$,
$|\cos\theta_{T}|$, where $\theta_{T}$ is the angle between
the thrust axes of the $B$~candidate and the rest of the event
in the CM frame, and $R_{2}$.
The variables $\Delta E$, $\Delta E^{ROE}$, and $m^{ROE}_{ES}$
are most effective at rejecting \BB\ background, especially
for events with two semileptonic decays, which have large missing energy. 
For continuum suppression, the event-shape variables 
$|\cos\theta_{T}|$ and $R_{2}$ are most useful.
The distribution for the background-discrimination variable $i$ is 
parameterized by a PDF $\eta^{j}_{i}$ for each component $j$ of the sample: 
$j = 1$ for signal, $j = 2$ for \BB\
background, and $j = 3$ for continuum background.
A likelihood is then computed for each sample component
as the product of the individual PDFs for that component:
${\cal L}^{j} = \prod_{i}\eta^{j}_{i}$. The ratio between the 
likelihood for the signal component and the sum of the signal
and background likelihoods then provides the final
discriminating variable, the likelihood ratio
${\cal L}_{R} = {\cal L}^{signal} / 
 ({\cal L}^{signal} + {\cal L}^{\BB} +  {\cal L}^{cont})$.

Using the simulation, the cut on ${\cal L}_{R}$ is optimized to
maximize the statistical significance of the signal.
This optimization is performed separately for 
$e^{+} e^{-}$ modes and $\mu^{+} \mu^{-}$ modes in the three $m(X_{s})$
regions $m(X_{s}) < 0.6$~\gevcc, 
$0.6 < m(X_{s}) < 1.1$~\gevcc, and 
$1.1 < m(X_{s}) < 1.8$~\gevcc, 
resulting in the cuts 
${\cal L}_{R} > 0.3$, $0.4$, and $0.9$ for the 
$e^{+} e^{-}$ modes and 
${\cal L}_{R} > 0.2$, $0.6$, and $0.9$ for the $\mu^{+} \mu^{-}$ modes.

%%% sample composition for mES > 5.2
After applying all selection criteria, a sample of
349 \BtoXsee\ and 222 \BtoXsmumu\ candidates remains.
According to the simulation, the background left
at this stage of the analysis consists
mostly of \BB\ events (86\% and 72\% of the total
background in the electron
and muon channels, respectively).

\section{Maximum likelihood fit}
\label{sec:Fit}
We perform an extended, unbinned maximum likelihood fit
to the $m_{ES}$ distribution 
in the region $m_{ES} > 5.2$~\gevcc\ to extract the signal yield as well as
the shape and yield of the combinatorial background.
The likelihood function ${\cal L}$ consists of four components:
\begin{equation}
{\cal L} = \frac{  e^{-(N_{sig} + N_{c-peak} + N_{h-peak} + N_{comb})} }{ N! }
           \prod_{ k = 1 }^{N} 
            [(N_{sig} + N_{c-peak}) {\cal P}^{c-peak}_{k} +
             N_{h-peak} {\cal P}^{h-peak}_{k} +
             N_{comb} {\cal P}^{comb}_{k}]
\end{equation}
where $N$ and $k$ denote the total number and index of candidate events,
respectively.
$N_{sig}$, $N_{c-peak}$, $N_{h-peak}$, and $N_{comb}$ represent the yields of 
the signal, charmonium peaking background, hadronic peaking background, 
and combinatorial background, respectively, with corresponding PDFs given by 
${\cal P}^{c-peak}_{k}$,
${\cal P}^{h-peak}_{k}$, and ${\cal P}^{comb}_{k}$.
The signal PDF is the same as that for the charmonium peaking background
since it is extracted from the charmonium-veto data sample.

The signal PDF ${\cal P}^{c-peak}_{k}$ is described by a Gaussian
for $\mu^+ \mu^-$ modes as well as for $e^+ e^-$ modes, since the 
Bremsstrahlung recovery and selection procedure for $e^+ e^-$ modes 
lead to a negligible radiative tail in the $m_{ES}$ distribution.
The Gaussian shape parameters are determined from fits to the sum of 
a Gaussian and Argus~\cite{ARGUS} function 
for the charmonium-veto data sample.
The fits yield signal $m_{ES}$ peak positions and resolutions of 
$m_{sig} = 5280.04 \pm 0.05$~\mevcc\ and 
$\sigma_{sig} = 2.80 \pm 0.05$~\mevcc\ for the $e^+ e^-$ modes,
and
$m_{sig} = 5280.05 \pm 0.07$~\mevcc\ and
$\sigma_{sig} = 2.61 \pm 0.06$~\mevcc\ for the $\mu^{+} \mu^{-}$ modes,
respectively.
In the simulation,
the Gaussian fit results for the $m_{ES}$ distributions for correctly
reconstructed signal are in agreement with the shape parameters
extracted from the fits to the charmonium-veto sample.
The signal yield $N_{sig}$ is a free parameter in the likelihood fit.
 
The charmonium peaking background is estimated from the simulation
to contribute $N_{c-peak}  = 0.6 \pm 0.3$ events in the electron modes and
$N_{c-peak} = 0.3 \pm 0.2$ events in the muon modes.
In the likelihood fit, $N_{c-peak}$ is fixed to these values. 

The size and shape of the hadronic peaking \BB\ background component arising 
from $B \to D^{(*)} n \pi$~$(n > 0)$ decays with misidentification 
of two charged pions as leptons
are derived directly from data by performing the analysis without the lepton 
identification requirements. Parameters for
the PDF ${\cal P}^{h-peak}_{k}$ are taken from a fit to the sum of a 
Gaussian and Argus function, with $m_{ES}$ peak positions and resolutions of
$m_{peak} = 5280.13 \pm 0.06$~\mevcc\ and
$\sigma_{peak} = 2.99 \pm 0.07$~\mevcc.
Taking the $\pi$ to $\ell$ misidentification rates into
account, the remaining hadronic peaking backgrounds 
are estimated to be $N_{h-peak} = 2.4 \pm 0.8$ events for 
the $\mu^{+} \mu^{-}$ modes and are
negligible for the $e^+ e^-$ modes. In the likelihood fit, $N_{h-peak}$ is fixed to
the estimated values.

The combinatorial background PDF, ${\cal P}^{comb}_{k}$, is given by an Argus
shape, which describes the combinatorial contribution from continuum events and 
\BB\ events. The Argus cutoff is determined by the 
beam energy in the $\Upsilon(4S)$ rest frame, $E_{beam} = 5.290$~\gev, 
whereas the Argus shape parameter and the yield $N_{comb}$ are free
parameters in the likelihood fit.

\section{Results}
\label{sec:Results}

\begin{figure}[!htb]
\begin{center}
\includegraphics[width=16cm]{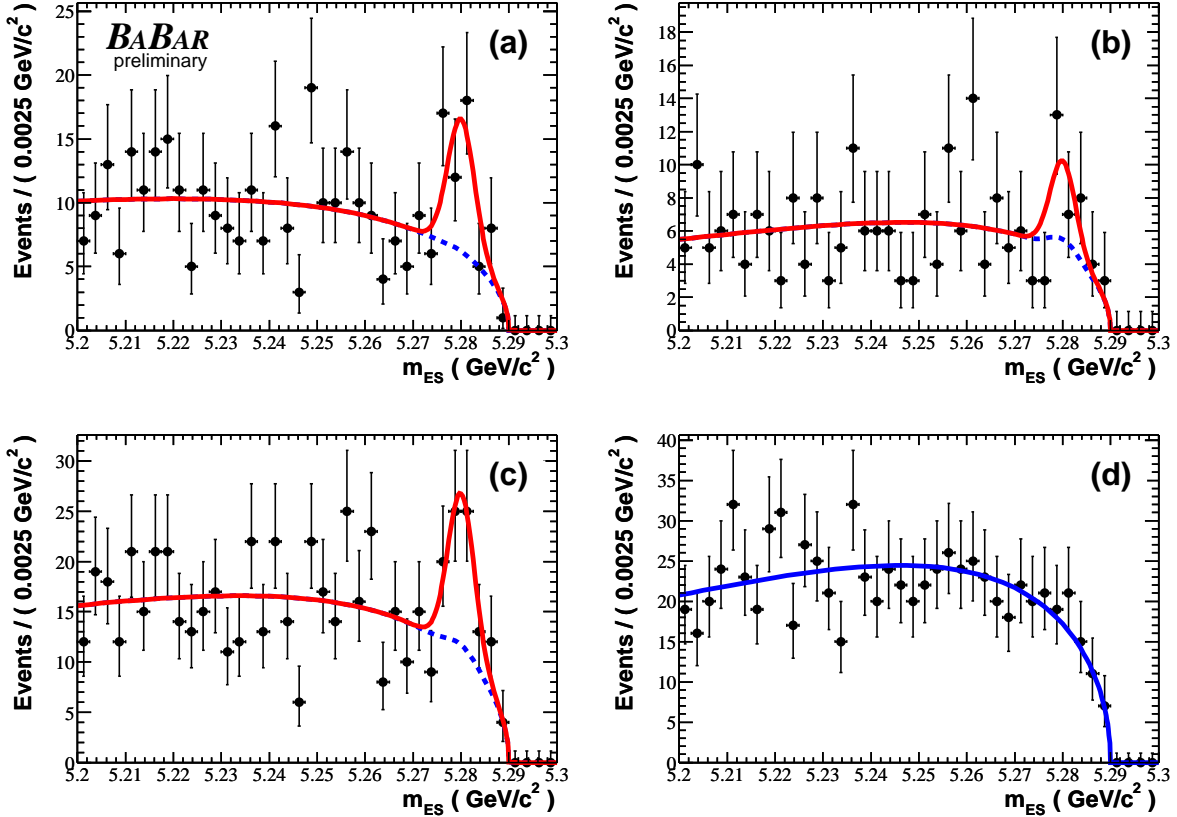}
\caption{Distributions of $m_{ES}$ for selected
(a) \BtoXsee, (b) \BtoXsmumu, (c) \BtoXsll\ ($\ell = e, \mu$), 
and (d) \BtoXsemu\ candidates.
The solid lines represent the result of the fits
and the dashed lines represent the background component
under the signal peaks.}
\label{fig:fitfig}
\end{center}
\end{figure}

Using the fit parameterizations described above, 
we fit the $m_{ES}$ distributions for the selected
\BtoXsee\ and \BtoXsmumu\ candidates separately and obtain the results shown in 
Figure~\ref{fig:fitfig}.
The fit results are summarized in Table~\ref{tab:fitresult}.
The statistical significance is
${\cal S} = \sqrt{2 (\ln{\cal L}_{max} - \ln{\cal L}^{0}_{max})}$,
where ${\cal L}_{max}$ represents the maximum likelihood for the
fit and ${\cal L}^{0}_{max}$ denotes the maximum likelihood for a
different fit when the signal yield is fixed at $N_{sig} = 0$.
The significance does not incorporate effects due to systematic
uncertainties in the signal shape or in the contribution from
peaking background. However, these have been found to be small.
The \BtoXsll\ signal yield presented in Table~\ref{tab:fitresult} is
the sum of the \BtoXsee\ and \BtoXsmumu\ signal yields.
A separate fit to the combined electron and muon channels
gives a very similar result.

Figure~\ref{fig:fitfig}(d) shows the distribution of $m_{ES}$ for \BtoXsemu\ 
candidates selected using the nominal selection criteria but
requiring that the two leptons have different flavor. 
This sample provides a cross-check for the parameterization
of the combinatorial background.
Figure~\ref{fig:fitfig}(d) shows an Argus fit to the $m_{ES}$ distribution 
and, as expected, there is no evidence for peaking background. 
The Argus shape and yield parameters for the combinatorial background
are free in the nominal fit to data.

\begin{figure}[!htb]
\begin{center}
\includegraphics[width=12cm]{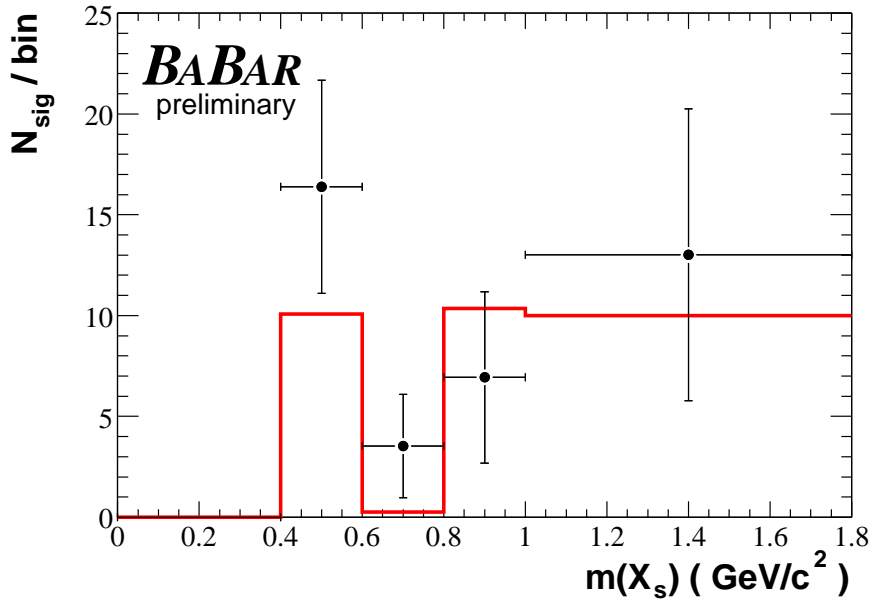}
\caption{Distribution of number of signal events as a function
 of hadronic mass for electron and muon channels combined
 for data (points) and Monte Carlo signal (histogram).
 The vertical error bars represent statistical uncertainties only.}
\label{fig:mXs}
\end{center}
\end{figure}

Figure~\ref{fig:mXs} shows the distribution of $m(X_{s})$ for electron 
and muon channels combined.
This distribution is obtained by performing the nominal likelihood fit
in separate $m(X_{s})$ regions.
Figure~\ref{fig:mXs} indicates that the observed signal receives contributions
from final states across a range of hadronic mass, including
hadronic systems with mass above that of the $K^\ast(892)$. 

\begin{table}[tb]
 \caption{Results of the fit to data: signal yield,
  charmonium and hadronic peaking backgrounds (fixed in the fit),
  combinatorial background yield, and statistical significance.}
 \begin{center}
 \begin{tabular}{lccccc}
 \hline\hline
  Sample & 
       $N_{sig}$       & $N_{c-peak}$ & $N_{h-peak}$ & $N_{comb}$ & Signif.\\
  \hline
  $X_s\: e^+ e^-$ &    
       $29.0 \pm ~8.3$ &   0.6   &  0.0      & $319.4 \pm 18.9$ &  4.0 \\
  $X_s\: \mu^+\mu^-$ & 
       $12.4 \pm ~6.2$ &   0.3   &  2.4      & $207.0 \pm 15.2$ &  2.2 \\
  $X_s\: \ell^+\ell^-$ &
       $41.4 \pm 10.3$ &   0.9   &  2.4      & $526.4 \pm 24.3$ &  4.6 \\
 \hline\hline
 \end{tabular}
 \label{tab:fitresult}
 \end{center}
\end{table}

The branching fraction ${\cal B}$ for the signal is calculated from
\begin{equation}
  {\cal B} = \frac{ N_{sig} }{ 2 N_{\BB}~ \epsilon },
\end{equation}
where $N_{\BB} = (88.9 \pm 1.0) \times 10^{6}$ is the number
of \BB\ pairs produced in $81.9$~\invfb\ and
$\epsilon$ is the signal efficiency.
The fitted signal yield $N_{sig}$ 
contains a contribution from misreconstructed \BtoXsll\ decays 
(cross-feed events) of $1.8 \pm 1.8$ events in the electron channel and 
$0.9 \pm 0.9$ events in the muon channel.
The central values are derived from fits to data in which the nominal 
fit function is changed to include a cross-feed PDF whose shape and 
normalization with respect to that for the correctly reconstructed \BtoXsll\ 
decays are taken from the simulation.
The corresponding uncertainties are estimated from Monte Carlo studies 
using a large number of data-sized experiments.
The signal efficiency $\epsilon$ is adjusted to reflect the
contribution from cross-feed events.

\section{Systematic uncertainties}
\label{sec:Systematics}

Systematic uncertainties are of two different
types: those that affect the extraction of the number of signal decays 
and those that affect the calculation of the branching fraction.
The systematic uncertainties are summarized in Table~\ref{tab_systematics}.

Uncertainties affecting the extraction of the signal yield
are evaluated by varying the signal Gaussian parameters (mean and width)
and the signal shape (using asymmetric signal shapes) within the 
constraints allowed by the charmonium veto data.
The small amount of peaking background from \BtoJpsiX\ and
\BtopsipX\ decays is varied according to the estimated number
of such decays: $0.6 \pm 0.3$ and $0.3 \pm 0.2$ in the
electron and muon channels, respectively.
The peaking background in the muon channel
due to \BtoXshh\ decays is varied according to the
estimated number of such events: $2.4 \pm 0.8$,
where the uncertainty is dominated by the uncertainty in the
$\pi$ to $\mu$ misidentification rates determined from
control samples in the data.
The shape of this background is also varied according to the
data sample selected without lepton identification.
The largest variations in the signal yield
are observed for changes in the signal
shape and in the amount of peaking background.

Uncertainties affecting the signal efficiency originate from
the detector modeling, from the simulation of signal decays,
and from the estimate of the number of $B$ mesons in the sample.
By far the dominant component is that due to the
simulation of signal decays, discussed in detail below.

The detector modeling uncertainty is sensitive to the
following uncertainties determined from the data:
the uncertainty in the tracking efficiency of 1.3\% per track
(2.0\% per charged pion); the uncertainty in the charged-particle
identification efficiency of 0.7\% per electron, 1.6\% per muon,
1.0\% per kaon and 2.0\% per pion; and the uncertainty in the reconstruction
efficiency of 10\% per \KS\ and 5\% per $\pi^0$.
The efficiency of the likelihood ratio cut to suppress combinatorial
background is checked with the charmonium-veto sample and the level
of discrepancy with the simulation is taken as the corresponding uncertainty.
The fraction of signal cross-feed included in the signal Gaussian is
studied with a parameterized Monte Carlo simulation technique and 
the uncertainty is set to
correspond to the R.M.S. obtained for a large number of
data-sized experiments (the uncertainty amounts
to a variation of $\pm 100\%$ of the number of signal cross-feed events
included in the Gaussian).

The dominant source of uncertainty arises from modeling signal decays.
Parameters of the Fermi motion model are varied in accordance with
measurements of hadronic moments in semileptonic $B$ decays~\cite{CLEO_moments}
and the photon spectrum in inclusive $B \to X_s\:\gamma$ decays~\cite{CLEO_bsgamma}.
The fraction of exclusive \BtoKll\ and \BtoKstarll\ decays
is varied according to theoretical uncertainties~\cite{Ali02}
and the corresponding systematic uncertainties are added linearly
(i.e., we assume the uncertainties in the branching fraction for
these two modes are fully correlated).
The transition point in $m(X_s)$ between pure $K^\ast \ell^+ \ell^-$
and nonresonant $X_s\:\ell^+\ell^-$ final states is varied
by $\pm 0.1$~\gevcc.

The nonresonant Monte Carlo event generator relies on JETSET to
fragment and hadronize the system consisting of a final
state $s$ quark and a spectator quark from the $B$ meson.
Since the signal efficiencies depend strongly on the
particle content of the final state, uncertainties in the
number of charged and neutral pions and in the 
number of charged and neutral kaons translate
into a significant uncertainty in the signal efficiency
(for $m(X_s) > 1.1$~\gevcc).

The ratio between the generator yield for decay modes containing
a \KS\ and that for modes containing a charged kaon
is varied according to $0.50 \pm 0.05$,
to allow for isospin violation in the decay chain.
The ratio between the generator yield for
decay modes containing one $\pi^0$ and that for modes
containing none is varied according to $1.0 \pm 0.5$.
The ratio between the generator yield for decay modes
into two-body hadronic systems and that into three-body hadronic
systems is varied according to $0.5 \pm 0.3$.
Uncertainties in the last two ratios are set by the
level of discrepancy between data and Monte Carlo
as measured in the \babar\ semi-inclusive
$B \to X_s\: \gamma$ analysis~\cite{BaBar_bsgamma}.
(Both $B \to X_s\:\gamma$ and present analyses use
the same hadronization model.)

The ten modes selected in this analysis only capture about
50\% of the full set of final states.
Approximately half of the missing modes is due to
final states with a \KL\ meson and their contribution can be determined
from the \KS\ modes.
However, we need to account for the uncertainty in the fraction of modes
with too many pions or kaons (two extra kaons may be
produced via $s \bar{s}$ popping),
as well as for modes with photons
that do not originate from $\pi^0$ decays but rather
from $\eta$, $\eta^\prime$, $\omega$, etc.
For final states with $m(X_s) > 1.1$~\gevcc, we vary
these different fractions by $\pm 50\%$.

\begin{table}[tb]
\caption{Summary of fractional systematic uncertainties (in \%).
 The uncertainties in extracting the signal are presented first and
 those related to the signal efficiency and \BB\ counting are
 presented second.}
\begin{center}
\begin{tabular}{lrr}
\hline \hline
Source    	        & $X_s\: e^+e^-$  & $X_s\: \mu^+\mu^-$   \\ \hline\hline
Signal shape            & $\pm 4.3$       & $\pm 2.5$    \\
Peaking background      & $\pm 1.0$       & $\pm 5.8$    \\ \hline\hline
~~~~Signal yield total  & $\pm 4.4$       & $\pm 6.3$    \\ \hline\hline
Tracking efficiency     & $\pm 5.0$       & $\pm 4.8$    \\
Lepton identification efficiency    & $\pm 1.3$       & $\pm 3.2$    \\
Kaon identification efficiency      & $\pm 0.8$       & $\pm 0.8$    \\
$\pi^\pm$ identification efficiency & $\pm 1.5$       & $\pm 1.3$    \\
\KS\ efficiency         & $\pm 2.2$       & $\pm 1.9$    \\
$\pi^0$ efficiency      & $\pm 0.8$       & $\pm 0.5$    \\
LR cut efficiency       & $\pm 3.7$       & $\pm 3.3$    \\
Cross-feed efficiency   & $\pm 6.7$       & $\pm 7.4$    \\ \hline
~~~~Detector model subtotal & $\pm 9.7$       & $\pm 10.2$   \\ \hline
Fermi motion model      & $^{+10.2}_{-4.0}$ &$^{+10.8}_{-3.6}$ \\
${\cal B}(\BtoKll)$     & $\pm 9.6$      & $\pm 12.6$    \\
${\cal B}(\BtoKstarll)$ & $\pm 6.2$       & $\pm 6.8$    \\
$K^\ast$--$X_s$ transition     & $\pm 3.3$       & $\pm 4.1$    \\
Hadronization           & $\pm 11.1$      & $\pm 7.5$    \\
Missing modes           & $\pm 10.6$      & $\pm 8.9$    \\ \hline
~~~~Signal model subtotal & $^{+24.5}_{-22.6}$ & $^{+25.4}_{-23.3}$ \\ \hline
Monte Carlo statistics  & $\pm 2.5$       & $\pm 3.0$    \\
\BB\ counting           & $\pm 1.1$       & $\pm 1.1$    \\
\hline\hline
~~~~Efficiency and $N_{\BB}$ total & $^{+26.5}_{-24.8}$ & $^{+27.6}_{-25.6}$   \\
\hline\hline
\end{tabular}
\end{center}
\label{tab_systematics}
\end{table}

Including systematic uncertainties, the measured branching fractions
for $m(\ell^{+} \ell^{-}) > 0.2$~\gevcc are
\begin{eqnarray}
 {\cal B}(\BtoXsee) & = & \left(6.6 \pm 1.9 ^{+1.9}_{-1.6}\right)
 \times 10^{-6}, \\
 {\cal B}(\BtoXsmumu) & = & \left(5.7 \pm 2.8 ^{+1.7}_{-1.4}\right)
 \times 10^{-6}, \\
 {\cal B}(\BtoXsll) & = & \left(6.3 \pm 1.6 ^{+1.8}_{-1.5}\right)
 \times 10^{-6},
\end{eqnarray}
where the first error is statistical and the second error is systematic.
The combined \BtoXsll\ branching fraction is the weighted average of
the branching fractions for the electron and muon channels,
where we assume the individual branching fractions to be equal 
for $m(\ell^{+} \ell^{-}) > 0.2$~\gevcc.
Table~\ref{tab_results} summarizes the results of the analysis
and lists both the statistical and systematic errors in the signal
yields, the signal efficiencies and the branching fractions.

\begin{table}
 \caption{Summary of results: signal yield,
  statistical significance, efficiency and branching fraction.
  In the case of the signal yield and the branching fraction,
  the first error is statistical and the second error is systematic.
  In the case of the signal efficiency, the first error corresponds
  to uncertainties in detector modeling, \BB\ counting, and Monte Carlo
  statistics, whereas the second error corresponds to the uncertainties
  in the signal model.}
 \baselineskip=28pt
 \begin{center}
 \begin{tabular}{lcccc}
  \hline \hline 
  Sample
    & $N_{sig}$ & Signif. & $\epsilon$ (\%) & ${\cal B}~(\times 10^{-6})$ \\
  \hline\hline
  $X_s\: e^+ e^-$
    & $29.0 \pm ~8.3 \pm 1.3$ & 4.0
    & $2.46 \pm 0.25 ^{+0.60}_{-0.56}$ & $6.6 \pm 1.9 ^{+1.9}_{-1.6}$ \\
  $X_s\: \mu^+\mu^-$
    & $12.4 \pm ~6.2 \pm 0.8$ & 2.2
    & $1.23 \pm 0.13 ^{+0.31}_{-0.29}$ & $5.7 \pm 2.8 ^{+1.7}_{-1.4}$ \\
  $X_s\: \ell^+\ell^-$
    & $41.4 \pm 10.3 \pm 1.5$ & 4.6
    &             ---                  & $6.3 \pm 1.6 ^{+1.8}_{-1.5}$ \\
  \hline
 \end{tabular}
  \label{tab_results}
 \end{center}
\end{table}

\section{Summary}
\label{sec:Summary}

Using a sample of $88.9 \times 10^{6}$
$\Upsilon(4S) \to B\overline{B}$ events,
we measure the branching fraction for the rare decay 
\BtoXsll, where $\ell = e$ or $\mu$,
with a sum over ten different hadronic states (with up to two pions).
For $m(\ell^{+} \ell^{-}) > 0.2$~\gevcc, we observe a signal of
$41 \pm 10(stat) \pm 2(syst)$ events
and obtain a branching fraction of
\begin{equation}
  {\cal B}(B \to X_{s}\:\ell^{+} \ell^{-}) = 
  \left(6.3 \pm 1.6(stat) ^{+1.8}_{-1.5}(syst)\right) \times 10^{-6},
\end{equation}
with a statistical significance of $4.6\:\sigma$.
This preliminary measurement agrees well with the recent prediction
by Ali {\em et al.}~\cite{Ali02}
and the measurement performed by the Belle Collaboration,
${\cal B}(B \to X_{s}\:\ell^{+} \ell^{-}) = 
 \left(6.1 \pm 1.4(stat) \pm^{1.4}_{1.1}(syst)\right) \times 10^{-6}$ 
for $m(\ell^{+} \ell^{-}) > 0.2$~\gevcc~\cite{Belle_sll}.

\pagebreak

\section*{Acknowledgments}
\label{sec:Acknowledgments}

% Specific acknowledgments for this paper; remove if not needed.
The authors wish to thank Gudrun~Hiller for her help.
We have also benefitted from discussions with Tobias Hurth.

% Standard acknowledgments paragraph; must always be included.
We are grateful for the 
extraordinary contributions of our \pep2\ colleagues in
achieving the excellent luminosity and machine conditions
that have made this work possible.
The success of this project also relies critically on the 
expertise and dedication of the computing organizations that 
support \babar.
The collaborating institutions wish to thank 
SLAC for its support and the kind hospitality extended to them. 
This work is supported by the
US Department of Energy
and National Science Foundation, the
Natural Sciences and Engineering Research Council (Canada),
Institute of High Energy Physics (China), the
Commissariat \`a l'Energie Atomique and
Institut National de Physique Nucl\'eaire et de Physique des Particules
(France), the
Bundesministerium f\"ur Bildung und Forschung and
Deutsche Forschungsgemeinschaft
(Germany), the
Istituto Nazionale di Fisica Nucleare (Italy),
the Foundation for Fundamental Research on Matter (The Netherlands),
the Research Council of Norway, the
Ministry of Science and Technology of the Russian Federation, and the
Particle Physics and Astronomy Research Council (United Kingdom). 
Individuals have received support from 
the A. P. Sloan Foundation, 
the Research Corporation,
and the Alexander von Humboldt Foundation.

\pagebreak

\end{document}